\documentstyle[12pt]{article}
\begin{document}
\renewcommand{\theequation}{\thesection.\arabic{equation}}
\renewcommand{\section}[1]{\addtocounter{section}{1}
\vspace{5mm} \par \noindent
  {\bf \thesection . #1}\setcounter{subsection}{0}
  \par
   \vspace{2mm} } 
\newcommand{\sectionsub}[1]{\addtocounter{section}{1}
\vspace{5mm} \par \noindent
  {\bf \thesection . #1}\setcounter{subsection}{0}\par}
\renewcommand{\subsection}[1]{\addtocounter{subsection}{1}
\vspace{2.5mm}\par\noindent {\em \thesubsection . #1}\par
 \vspace{0.5mm} }
\renewcommand{\thebibliography}[1]{ {\vspace{5mm}\par \noindent{\bf
References}\par \vspace{2mm}}
\list
 {\arabic{enumi}.}{\settowidth\labelwidth{[#1]}\leftmargin\labelwidth
 \advance\leftmargin\labelsep\addtolength{\topsep}{-4em}
 \usecounter{enumi}}
 \def\newblock{\hskip .11em plus .33em minus .07em}
 \sloppy\clubpenalty4000\widowpenalty4000
 \sfcode`\.=1000\relax \setlength{\itemsep}{-0.4em} }
\newcommand\rf[1]{(\ref{#1})}
\def\nn{\nonumber}
\newcommand{\sect}[1]{\setcounter{equation}{0} \section{#1}}
\renewcommand{\theequation}{\thesection .\arabic{equation}}
\newcommand{\ft}[2]{{\textstyle\frac{#1}{#2}}}
\newcommand{\Zint}{\mathbb{Z}}
\newcommand{\Real}{\mathbb{R}}
\newcommand{\be}{\begin{equation}}
\newcommand{\ee}{\end{equation}}
\newcommand{\ba}{\begin{eqnarray}}
\newcommand{\ea}{\end{eqnarray}}
\newcommand{\ensuremath}[1]{#1}
\newcommand{\eV}{\rm eV }
\newcommand{\TeV}{\rm TeV }
\newcommand{\GeV}{\rm GeV }
\newcommand{\MeV}{\rm MeV }
\newcommand{\Kt}{\ensuremath{K_3}}
\newcommand{\VKt}{\ensuremath{V_{K_3}}}
\newcommand{\lh}{\ensuremath{l_{\rm H}}}
\newcommand{\lm}{\ensuremath{l_{\rm M}}}
\newcommand{\lii}{\ensuremath{l_{\rm II}}}
\newcommand{\li}{\ensuremath{l_{\rm I}}}
\newcommand{\lp}{\ensuremath{l_{\rm P}}}
\newcommand{\mpl}{\ensuremath{M_{\rm P}}}
\newcommand{\mh}{\ensuremath{M_{\rm H}}}
\newcommand{\mii}{\ensuremath{M_{\rm II}}}
\newcommand{\mi}{\ensuremath{M_{\rm I}}}
\newcommand{\mgut}{\ensuremath{m_{\rm GUT}}}
\newcommand{\gh}{\ensuremath{\lambda_{\rm H}}}
\newcommand{\gii}{\ensuremath{\lambda_{\rm II}}}
\newcommand{\gi}{\ensuremath{\lambda_{\rm I}}}
\newcommand{\gip}{\ensuremath{\lambda_{\rm I'}}}
\newcommand{\gym}{\ensuremath{g_{\rm YM}}}
\newcommand{\giis}{\ensuremath{\lambda_{\rm 6IIA}}}
\newcommand{\ghs}{\ensuremath{\lambda_{\rm 6H}}}
\newcommand{\giia}{\ensuremath{\lambda_{\rm IIA}}}
\newcommand{\giib}{\ensuremath{\lambda_{\rm IIB}}}
\newcommand{\giibs}{\ensuremath{\lambda_{\rm 6IIB}}}
\newcommand{\irrep}[1]{\ensuremath{\boldsymbol{#1}}}
\newcommand{\ie}{\hbox{\it i.e.}\ }
\newcommand{\Tr}{{\rm Tr}\ }
\def\simlt{\stackrel{<}{{}_\sim}}
\def\simgt{\stackrel{>}{{}_\sim}}
\thispagestyle{empty}

\begin{flushright}
{\tt hep-ph/9906480}\\
CPHT-PC720.0699\\
\end{flushright}

\begin{center}

\vspace{2cm}

{\large\bf Large Dimensions \\ and String Physics at a TeV
\footnote{Research supported in part by the EEC under the TMR contract
ERBFMRX-CT96-0090.}}\\

\vspace{1.4cm}

{\sc I. Antoniadis and B. Pioline}\\

\vspace{1.3cm}

{\em Centre de Physique Th\'eorique (CNRS UMR 7644)} \\
{\em \'Ecole Polytechnique} \\
{\em F-91128 Palaiseau Cedex, France}\\
\vspace{1cm}

\centerline{\bf Abstract}
\vspace{- 2 mm}  \end{center}
\begin{quote}\small
Large extra dimensions, of size of order of TeV$^{-1}\simeq 10^{-16}$ cm,
arise naturally in the context of supersymmetry breaking in string theory,
while strings at a TeV scale offer a solution to the gauge hierarchy problem,
as an alternative to softly broken supersymmetry or technicolor.
In this short review,
\footnote{
talk presented by the first author at the conferences 
``Fundamental interactions: from symmetries to
black holes'' in honor of Francois Englert, Brussels, March 25-27, 1999
and ``Beyond the Desert 99'', Castle Ringberg, Tegernsee, Germany, June 6-12,
1999.}
we present consistent perturbative 
realizations of string theories with large volume compactifications and
low string tensions, and discuss their main physical implications. 
\end{quote}
\vfill

\eject
\baselineskip18pt
\addtocounter{section}{1}
\par \noindent
{\bf \thesection . Introduction}
\par \vspace{2mm} 
\noindent  
Since the discovery that superstring theory was providing an ultraviolet
regulator of quantum gravity, it has been customary to associate it with
sub-Planckian physics, far out of reach of laboratory experiments.
This expectation is natural in the conventional supersymmetric
unification scenario at $\mgut\simeq 3\times10^{16}$ GeV, and met 
in the phenomenologically most promising weakly coupled heterotic
string theory, which gives a qualitative description thereof. 
Despite this success, there are some physical motivations 
suggesting that large volume compactifications may be relevant for
physics. One comes from the quantitative description of the gauge coupling
unification which is off by roughly two orders of magnitude. Another
results from the problem of supersymmetry breaking in string theory 
that requires a compactification scale of the order of a few TeV \cite{ia}. 
This is one of the very few general 
predictions of (perturbative) string theory, 
which relates the supersymmetry breaking scale to the
size of internal compact dimensions \cite{ablt}. 

Large volume compactifications imply that the heterotic string is strongly 
coupled and is described by some dual weakly coupled theory, namely
type I (in general type I$^\prime$), or type II (IIA or IIB) \cite{sen}. 
A general 
property of these theories is that gauge interactions are localized on
$p$-branes with $p\le 9$ and therefore gauge particles and gravity propagate
in different spacetime dimensions \cite{pol}. In this new context,
the four-dimensional (4d) Planck length $\lp\equiv\mpl^{-1}\simeq 10^{-33}$ cm
is no longer thought of as a fundamental microscopic
scale, but as a derived quantity from the string scale 
$l_s\equiv M_s^{-1}$, the
compactification parameters and the string coupling $\lambda_s$ \cite{w}.
The fundamental string scale can therefore be anywhere above 
the TeV scale \cite{l}.
In particular, if it were at the TeV, the gauge hierarchy problem would be
nullified without the need for low energy supersymmetry or technicolor
\cite{add,aadd,ab}, that protect the Higgs mass \cite{higgs} 
against quadratically divergent corrections. The price to pay is to introduce
hierarchically large transverse dimensions felt only by gravity (type
I/I$^\prime$) \cite{aadd,st}, or
an extremely small string coupling (type II) \cite{ap}, and in 
our present state of understanding to renounce to the 
conventional gauge unification.
The gain is a new prospect on the gauge hierarchy,
supersymmetry breaking and possibly cosmological
constant problems, and most excitingly potential experimental
signatures at next-round colliders. In the following, we will
review the possibilities for lowering the fundamental scale in 
various string frameworks, and discuss their main experimental consequences.

Unification of gauge interactions and gravity would be achieved
upon finding a particular string vacuum that reproduces the particle spectrum,
masses and couplings of the Standard Model, together
with the correct Newton constant and quasi-vanishing cosmological constant.
By lack of an acceptable candidate, we will restrict ourselves to the 
requirement that the correct Planck mass $\mpl=1.2\times 10^{19}$ GeV 
and gauge
coupling constant $\gym\simeq 1/5$, describing 
the gravitational and 
electrostatic interactions in units of $\hbar$ and $c$
\begin{equation}
E_{grav}=-\frac{M_1 M_2}{\mpl^2 r}\ ,\qquad
E_{elec}=\frac{\gym^2 Q_1 Q_2}{r}\ ,
\end{equation}
respectively, be reproduced. 
In particular, we identify the three gauge couplings
assuming standard perturbative unification.

\setcounter{equation}{0}
\section{TeV$^{-1}$ dimensions in heterotic string}
\noindent 
In heterotic string, gauge and gravitational interactions 
appear at the same (tree)
level of string perturbation theory corresponding to spherical worldsheets.
The 10d effective action is given on dimensional grounds by
\be
S=\int [d^{10}x] \left\{
\frac{1}{\gh^2 \lh^8} {\cal R} + \frac{1}{\gh^2 \lh^6} F^2 + \dots\right\}\, ,
\label{actionh}
\ee
in a self-explanatory notation.
Upon compactification on a six-manifold of volume $V_6$,
we read off the Planck mass and gauge coupling:
\be
\frac{1}{\lp^2}=\frac{V_6}{\gh^2 \lh^8}\ ,\qquad
\frac{1}{\gym^2}=\frac{V_6}{\gh^2 \lh^6}
\label{treeh}
\ee
so that at tree level
\be
\label{treehc}
\mh=\gym\mpl\ ,\qquad \gh=\gym \sqrt{V_6/\lh^6}\, .
\ee
Given that $\gym\sim 1$, the Planck and string scales
appear to be intimately tied, thus justifying the afore-mentioned
prejudice. More precisely, one finds $\mh\simeq 10^{18}$ GeV, 
which is a factor of
30 larger than the unification scale $\mgut$. 
This may be considered either as a
success in comparison of the hierarchy between the weak scale and the 
unification scale, or as a mild call for going beyond the weak coupling
regime \cite{w}. This identification is of course valid when the
string coupling is small $\gh<1$, which implies that the compactification volume
should be roughly of string size $V_6\sim\lh^6$.\footnote{We
restrict ourselves to compactification volumes larger than the string length since
for smaller volumes there are light winding modes that can be traded for
ordinary Kaluza-Klein states by use of T-duality.}

By introducing a large TeV dimension, needed to break
supersymmetry by compactification, the string coupling $\gh$ becomes huge,
invalidating the perturbative description. From the 4d point
of view, the problem arises due to the infinite massive tower of 
Kaluza-Klein (KK)
excitations that are produced at energies above the compactification scale and
contribute to physical processes: 
\be
M^2=M_0^2+{m^2\over R^2}\ ;\qquad m=0,\pm 1,\dots\, ,
\label{KK}
\ee
where $M_0$ is a 5d mass and $R$ is the radius of an extra (fifth) dimension.
In this context,
a possible way out consists of imposing a set of conditions to the low
energy theory that prevent the effective couplings to diverge \cite{ia}. 
An example
of such conditions is that KK modes should 
be organized into multiplets of $N=4$
supersymmetry, containing for each spin-1 particle 4 Weyl fermions 
and 6 scalars with the
same quantum numbers, so that their 
contribution to beta-functions vanishes for every
$m\neq 0$ and gauge couplings remain finite.

The problem of strong coupling can now be addressed using the recent results
on string dualities. For instance, in the minimal case of 
one large dimension, the
dual weakly coupled description is provided by type IIB string theory, 
the tension of which 
appears as a non-perturbative threshold below the heterotic scale \cite{ap}:
$\lii\sim\gym\sqrt{R\lh}$, so that $\mii$ is at intermediate energies
of order $10^{11}$ GeV when $R\sim 1$ TeV$^{-1}$. At energies above the TeV but
below the IIB string scale $\mii$, the effective higher-dimensional theory is
described by a non-trivial infrared fixed point of the renormalization 
group \cite{scfp},
which encodes the conditions needed to be imposed in the low energy
couplings in order to ensure a smooth ultraviolet (UV) behavior, 
in the absence of 
gravity. These conditions generalize the requirement of $N=4$ supersymmetry 
for the
KK excitations, that keep gauge couplings well behaved, to the Yukawa and other
couplings of the theory. A generic property of these models is that chiral matter is
localized at particular points of the large internal dimensions. As a result, quarks
and leptons have no KK TeV excitations, 
which is welcome also for phenomenological
reasons in order to avoid fast proton decay \cite{ia,abe}. It is  
remarkable that the
main features of these models were captured already in the context of 
the heterotic string despite its strong coupling.

\setcounter{equation}{0}
\section{Realization of TeV strings}
\subsection{Type I strings}
In ten dimensions, the strongly coupled $SO(32)$ heterotic string is 
described by the type I string, which upon T-duality to 
type I$^\prime$ is actually
equivalent to the Horava-Witten M-theory.
Type I/I$^\prime$ is a theory of closed and open unoriented strings. 
Closed strings describe
gravity, while gauge interactions are described by open strings whose ends are
confined to propagate on D-branes. As a result, gauge and 
gravitational interactions
appear at different order in perturbation theory and the effective action reads
\begin{equation}
S=\int [d^{10}x] \frac{1}{\gi^2 \li^8} {\cal R} + 
\int [d^{p+1}x] \frac{1}{\gi^2 \li^{p-3}} F^2 + \dots\, ,
\end{equation}
where the $1/\gi$ factor in the gauge kinetic terms confined on a $p$-brane
corresponds to the disk diagram. 

Upon compactification to four dimensions, the
Planck length and gauge couplings are given at leading order by
\begin{equation}
\frac{1}{\lp^2}=\frac{V_\parallel V_\perp}{\gi^2 \li^8}\ ,\qquad
\frac{1}{\gym^2}=\frac{V_\parallel}{\gi \li^{p-3}}\, ,
\end{equation}
where $V_\parallel$ ($V_\perp$) denotes the compactification volume 
longitudinal
(transverse) to the $p$-brane. In this case, the requirement of weak coupling
$\gi<1$ implies that the size of the longitudinal space must be of order of the
string length ($V_\parallel\sim\li^{p-3}$), while the transverse volume
$V_\perp$ remains unrestricted. One thus has
\begin{equation}
\lp^2=\frac{\gym^4 v_\parallel \li^{2+n}}{R_\perp^n}\ ,\qquad
\gi=\gym^2 v_\parallel\, ,
\label{treei}
\end{equation}
to be compared with the heterotic relations (\ref{treehc}). Here, 
$v_\parallel\simgt 1$ is the longitudinal volume in string units, 
and we assumed
an isotropic transverse space of $n=9-p$ compact 
dimensions of radius $R_\perp$. 

Taking the type I string scale to be at the TeV, one finds a size for the
transverse dimensions varying from $10^8$ km, .1 mm, down 
to .1 fermi for $n=1,2$, or 6 large dimensions. 
The case $n=1$ is obviously experimentally excluded, but,
as we shall discuss in Section 4, all the other possibilities 
are consistent with observations, although barely in the
$n=2$ case \cite{add2}. In particular, sub-millimeter transverse
directions are compatible with 
the present constraints from short-distance gravity measurements \cite{add}.
An important property of these models is that gravity becomes strong
at the TeV, although the string coupling $\gi$ remains weak. In fact,
the first relation of eq. (\ref{treei}) can be understood as a
consequence of the $(4+n)$-dimensional Gauss law for gravity, with
$\gym^4\li^{2+n} v_\parallel$ the Newton constant in $4+n$ dimensions.
\subsection{Type IIA strings}
Upon compactification to 6 dimensions or lower, the heterotic string
admits another dual description in terms of type IIA string theory
compactified on a Calabi-Yau manifold. For simplicity, we shall
restrict ourselves to compactifications on $\Kt \times
T^2$, yielding $N=4$ supersymmetry, even though 
more phenomenological models with $N=1$ supersymmetry 
would require F-theory on a Calabi-Yau
four-fold, which is poorly understood at present.
In contrast to  heterotic and type I strings,
non-abelian gauge symmetries in type IIA models arise non-perturbatively 
(even though at arbitrarily weak coupling)
in singular compactifications, where the massless gauge bosons
are provided by D2-branes wrapped around non-trivial vanishing 2-cycles.
The resulting gauge interactions are localized on $\Kt$
(similar to a Neveu-Schwarz five-brane), while matter multiplets
would arise from further singularities, localized completely on the 6d internal
space.

It follows that gauge kinetic terms are independent of the string coupling
$\giia$ but given instead by the size $v_{T^2}$ of the two-torus $T^2$
in string units, so that
\begin{equation}
\frac{1}{\gym^2}=v_{T^2} ,\quad
\frac{1}{\lp^2}=\frac{v_{T_2} v_{{\Kt}} }{\giia^2 \lii^2 }\ .
\label{relii}
\end{equation}
The volume of $T^2$ should therefore be of order $\lii^2$,
to reproduce $\gym\sim 1$, while the Planck scale is expressed by
\begin{equation}
\lp=\gym ~\giia~ v_{\Kt}^{-1/2}~  \lii\, .
\end{equation}
In contrast to the type I relation (\ref{treei}) where only the volume of the
internal six-manifold was appearing, we now have the freedom to use both the
string coupling and the volume of $\Kt$ 
to separate the Planck mass at $10^{19}$ GeV from a string scale at 1 TeV. 
In particular, we can choose a string-size internal manifold, and 
have an ultra-weak coupling $\gii=10^{-14}$ to account for the
hierarchy between the weak scale and the Planck scale \cite{ap}.
\subsection{Type IIB strings}
Finally, we may also consider type IIB constructions,
in which gauge symmetries still arise from vanishing
2-cycles of $\Kt$, but take the form of tensionless
strings in 6 dimensions, given by D3-branes wrapped on the 
vanishing cycles. Only after further compactification does
this theory reduce to 
a standard gauge symmetry, whose coupling involves the shape $u_{T^2}$ rather
than the volume of the torus $T^2$. 
\be
\frac{1}{\gym^2}=u_{T^2}\ ,\quad
\frac{1}{\lp^2}=\frac{v_{T^2} v_{\Kt}  }{\giib^2\lii^2}\ .
\label{reliib}
\ee
In the case of a rectangular torus, 
its shape is given by the ratio of its two radii $u_{T^2}=R_1/R_2$.

Comparing with eq. (\ref{relii}), it is clear that the situation in type IIB is
the same as type IIA, unless the size of $T^2$ is much larger than the string
length. Since $T^2$ is felt by gauge interactions, its size cannot be larger than
TeV$^{-1}$ implying that the type IIB string scale should be much larger than TeV.
Considering a rectangular torus of radii $R$ and $\gym^2 R$, one finds
\be
\lp= \gym^{-1}~ \giib~ v_{\Kt}^{-1/2}~ \frac{\lii^2}{R}\, ,
\ee
showing that the largest value for the string tension, when $R\sim{\rm TeV}^{-1}$,
is an intermediate scale $10^{11}$ GeV with $v_{\Kt}\sim\giib\sim 1$. This is
precisely the case that describes the heterotic string with one TeV dimension,
which we discussed is Section 2. It is the only example of longitudinal dimensions
larger than the string length in a weakly coupled theory. In the energy range 
between the KK scale $1/R$ and the string scale $1/\lii$, one has an effective 6d
theory without gravity at a non-trivial superconformal 
fixed point described by a
tensionless string \cite{scfp}. 
\subsection{Are there more possibilities ?}
As we mentioned, the type I/I$^\prime$ and type IIA/IIB theories
can be seen as dual descriptions of the heterotic string at strong coupling.
Somewhat surprisingly, it turns out that all TeV scale
string models that we have discussed can be recovered as 
different strongly coupled decompactification limits
of the heterotic string, assuming only one large scale
in addition to the Planck-size heterotic tension.
More precisely, we consider the heterotic string compactified on a
six-torus with $k$ large dimensions of radius $R\gg\lh$
and $6-k$ string-size dimensions. 
Since the string coupling is strong, the relations (\ref{treeh})
strictly hold  only for cases with maximal supersymmetry,
so that tree-level gravitational and gauge kinetic terms (\ref{treeh}) are 
not corrected, and it is possible though unlikely that we may
overlook new possibilities in restricting to this simple case.

Applying the standard duality map, it is quite easy to show that
the type I$^\prime$ theory with $n$ transverse dimensions offers a weakly coupled
dual description for the heterotic string with $k=4,5,6$ large dimensions.
$k=4$ is described by $n=2$, $k=6$ (for $SO(32)$ gauge group) 
is described by $n=6$, while
for $n=5$ one finds a type I$^\prime$ model with 5 large transverse dimensions
and one extra-large. The type II theory on the other hand provides a weakly
coupled description for $k=1,2,3,4$ and $k=6$ (for $E_8\times E_8$).
In particular, $k=1$ yields the type IIB model with string tension at
intermediate energies that we discussed in Sections 2 and 3.3; $k=2$ is
described by the type IIA model with infinitesimal coupling, while for $k=3$
the four (transverse) $\Kt$ directions should be extra large.

\setcounter{equation}{0}
\section{Experimental predictions}
The main predictions of string theories with large volume compactifications
and/or low fundamental scale discussed above follow from the existence of (i)
large {\it longitudinal} dimensions felt by gauge interactions,
(ii) extra large {\it transverse} transverse dimensions felt only 
by gravity that becomes
strong at low energies, and (iii) strings with low tension. In table
{\ref{theories}, we summarize how these possibilities can be realized in
various weakly coupled string theories.
\begin{figure}
\begin{center}
\hspace*{-1.4cm}
\begin{tabular}{|c|c|c|c|c|c|}
\hline
Theories & $\parallel$ TeV$^{-1}$ dims & $\perp$ dims & strong gravity & 
string scale \\
\hline
type I/I$^\prime$& $6-n$ & $n\ge 2$ (mm - fm)  & TeV            & TeV \\ 
type IIA & 2     & TeV$^{-1}$          & $10^{19} $ GeV & TeV \\
         & $6-n$ & $2\le n\le 4$ (mm - fm) & TeV            & TeV \\
type IIB & 2     & $10^{11}$ GeV       & $10^{11}$ GeV  & $10^{11}$ GeV \\
\hline
\end{tabular}
\end{center}
\caption{Realizations of large dimensions and/or low string scale.\label{theories}}
\end{figure}
\subsection{Longitudinal dimensions}
They exist generically in all realizations of table {\ref{theories}, 
with the exception of type I TeV strings with six transverse dimensions 
in the fermi region.
Their main implication is the existence of KK excitations (\ref{KK}) for all
Standard Model gauge bosons and possibly the Higgs \cite{higgs}. 
They couple to quarks and leptons which are localized in the compact space, 
and generate at low
energies four-fermion and higher dimensional 
effective operators. The current limits
on their size arise then from the bounds of compositeness or from
other indirect effects, such as in the Fermi constant and LEP2 data, 
and lie in the
range of a couple of TeV \cite{abe,limits}, 
if the string scale is not far from the
compactification scale. 
Otherwise, for more than two longitudinal dimensions, the
limits are much higher because the sum over KK modes is power-like divergent
\cite{abe}. This is consistent with the type IIB realization in table
{\ref{theories}.

The most exciting possibility is of course their discovery through direct
production of KK excitations, for instance in hadron colliders such as 
the Tevatron and LHC, via Drell-Yan  processes \cite{abq}. 
The corresponding KK resonances are
narrow with a width-to-mass ratio $\Gamma/M\sim\gym^2\sim$ 
a few per
cent, and thus the typical expected signal is the production of a 
double resonance
corresponding the first KK mode of the photon and $Z$, 
very nearly spaced one
from the other. On the other hand, the 
non-observation of deviations from the Standard
Model prediction of the total number of lepton pairs at LHC would 
translate into a lower bound of about 7 and 9 TeV, for 
one and two large dimensions, respectively.
\vskip -0.1cm
\subsection{Supersymmetry breaking by compactification}
Large TeV$^{-1}$ dimensions can be used to break supersymmetry by
compactification. This breaking is realized through boundary conditions and is
similar to the effects of finite temperature with the identification $T\equiv
R^{-1}$. It follows that the breaking is extremely soft and insensitive to the
UV physics above the compactification scale. The summation over the KK
excitations amounts to inserting the Boltzmann factors $e^{-E/T}$ to all
thermodynamic quantities --or equivalently to the soft breaking terms--  that
suppress exponentially their UV behavior \cite{ia,add1}. This is in
contrast to the behavior of supersymmetric couplings that generally blow up, 
unless special conditions are imposed.

The extreme softness of supersymmetry breaking by compactification implies
a particular spectroscopy of superparticles that differs drastically 
from other
scenarios \cite{ia,adpq}. 
In the simplest case, supersymmetry breaking generates a
universal tree-level mass for gauginos, while scalar masses vanish
at tree-level. The latter
are insensitive to the UV cutoff at one loop, and thus squarks and leptons are
naturally an order of magnitude lighter than gauginos. On the other hand, 
if the Higgs scalar lives in the bulk of the extra (TeV) dimension(s), a heavy
higgsino mass is automatically generated and there is no $\mu$-problem. 
\vskip -0.1cm
\subsection{Transverse dimensions and low scale quantum gravity}
They exist generically in all type I/I$^\prime$ realizations of TeV strings and
their size can be as large as a millimeter, 
which is the shortest distance to which 
gravity has been directly tested experimentally \cite{price}. 
The strongest bounds obviously 
apply for the case of 2 transverse (sub)millimeter dimensions and 
come from astrophysics and cosmology \cite{add2}. Indeed, graviton emission 
during supernovae cooling 
restricts the 6d Planck scale to be larger than about 50 TeV,
implying $M_I\simgt 7$ TeV, while the graviton decay contribution 
to the cosmic
diffuse gamma radiation gives even 
stronger bounds of about 110 TeV and 15 TeV for
the two scales, respectively.

The main experimental signal in particle accelerators is graviton 
emission into the higher-dimensional 
bulk, leading to jets and missing energy events \cite{aadd}.
LHC will be sensitive to higher-dimensional gravity scales in the 
range of 3 to 5 TeV,
when the number of transverse dimensions varies from six at the  sub-fermi 
to two
at the sub-millimeter region, where the effect becomes stronger 
\cite{add2,lowgrav}. 
When the available energy becomes higher than the
gravity scale, gravitational interactions are strong and particle
colliders become the best probes for quantum gravity.
\vskip -0.1cm
\subsection{Low scale strings}
The main experimental signal of low scale strings is 
the production of higher-spin
Regge excitations for all Standard Model particles, with 
same quantum numbers and
mass-squared increasing linearly with spin. For instance, 
the excitations of the
gluon could show up as a series of peaks in jet production at LHC. However, the
corresponding resonances might be very narrow, with a width-to-mass ratio
$\Gamma/M\sim\gi^2$ of a few per thousand 
if $v_\parallel\sim 1$ in eq.
(\ref{treei}), and thus difficult to detect. On the other hand, in the 
type IIA realization of TeV strings of table \ref{theories} 
using an infinitesimal string
coupling $\giia\sim 10^{-14}$, string interactions are extremely suppressed and
there are no observable effects other than KK excitations of 
gauge particles, all
the way up to 4d Planckian energies.
\vskip -0.1cm
\subsection{Physics with large dimensions}
The possibility of large dimensions can be used to address many physical 
problems of
high energy physics from a new perspective. In particular, longitudinal (TeV)
dimensions can be used to break supersymmetry as in their original motivation
\cite{ia} that we discussed above, or to lower the unification scale \cite{gut}
using power law running \cite{tv}. On the other hand, extra large transverse
dimensions can be used to suppress proton decay by gauging baryon number 
in the bulk
\cite{st}, or to generate small neutrino masses and 
oscillations by identifying for
instance the right-handed neutrino with a bulk fermion \cite{nus}.
\vskip -0.1cm
\subsection{Gravity modification in the (sub)millimeter range}
Another category of predictions consists in
modifications of gravitation at short distances. These
deviations  could be measured in
experiments which test gravity in the sub-millimeter range \cite{price}.
There are two classes of such predictions \cite{ia2}:\hfil\\ 
(i) Deviations from the Newton law $1/r^2$ behavior to $1/r^{2+n}$, for $n$
extra large transverse dimensions, which can be observable for $n=2$ 
dimensions of sub-millimeter size.
This case is particularly attractive on theoretical grounds 
because of the logarithmic
sensitivity of Standard Model couplings on the size of transverse space 
\cite{ab},
but also for phenomenological reasons since 
the effects in particle colliders are
maximally enhanced. Notice also the coincidence of this scale with 
the possible
value of the cosmological constant in the universe that recent observations 
seem to
support.\hfil\\
(ii) New scalar forces in the sub-millimeter range, 
motivated by the problem of supersymmetry
breaking, and mediated by light scalar fields with masses 
\be
{m_{susy}^2\over\mpl}\simeq 10^{-4} - 10^{-2}\ {\rm eV} \, ,
\label{msusy}
\ee
for a supersymmetry breaking scale $m_{susy}\simeq 1-10$ TeV. 
These correspond to
Compton wavelengths in the range of 1 mm to 10 $\mu$m.
$m_{susy}$ can be either the KK scale $1/R$ if supersymmetry is broken by
compactification, or the string scale if it is broken ``maximally" on our
world-brane. Moreover, the scalar mediating the force is the radius modulus
$\ln R$ (in Planck units), with $R$ the radius of the longitudinal or 
transverse
dimension(s), respectively. Indeed, in the former case, this is due to the behavior of the vacuum energy density $\Lambda \sim 1/R^4$ for large $R$, up to 
logarithmic corrections
\cite{add1}. In the latter case, supersymmetry is broken primarily 
on the
brane only, and thus 
its transmission to the bulk is gravitationally suppressed, leading
to masses (\ref{msusy}) \cite{aadd}. The coupling of these light 
scalars to nuclei
can be computed since it arises dominantly through the radius dependence of
$\Lambda_{\rm QCD}$, or equivalently of the QCD gauge coupling. 
In the former case,
it is roughly $1/3$ $\times$ gravity, while in the latter 
it is again comparable to
gravity in theories with logarithmic sensitivity on the size of 
transverse space.
The resulting forces can therefore be within reach of upcoming experiments.


\end{document}